\documentclass[conference,10pt]{IEEEtran}
\IEEEoverridecommandlockouts
\usepackage{cite}
\usepackage{amsmath,amssymb,amsfonts}
\usepackage{algorithmic}
\usepackage{graphicx}
\usepackage{array}
\usepackage{multirow}
\usepackage{tabularx}
\usepackage{textcomp}
\usepackage{xcolor}
\def\BibTeX{{\rm B\kern-.05em{\sc i\kern-.025em b}\kern-.08em
    T\kern-.1667em\lower.7ex\hbox{E}\kern-.125emX}}

\usepackage{tikz}
\usetikzlibrary{patterns}
\usepackage{pgfplots}
\usetikzlibrary{plotmarks}
\usetikzlibrary{pgfplots.fillbetween}
\tikzstyle{block1} = [rectangle, text centered, draw=black, rounded corners]
\tikzstyle{block2} = [rectangle, text centered, draw=black, minimum width=0.5cm, minimum height=0.5cm]
\tikzstyle{block3} = [rectangle, text centered, draw=white, text width=1.5cm]
\tikzstyle{block4} = [rectangle, text centered, draw=white, text width=0.5cm]
\tikzstyle{block5} = [rectangle, text centered, draw=white, text width=0.4cm]
\tikzstyle{arrow} = [thick,->,>=stealth]
\tikzstyle{sum} = [circle, draw=black, very thick, minimum size=1mm]

\begin{document}

\title{Parallel APSM for Fast and Adaptive Digital SIC in Full-Duplex Transceivers with Nonlinearity\\
\thanks{Currently, O. Taghizadeh is with Lenovo Deutschland GmbH. His contributions were made when he was with the Network Information Theory Group, Technische Universität Berlin.}
}

\author{\IEEEauthorblockN{M. Hossein Attar, Omid Taghizadeh, Kaxin Chang, Ramez Askar, Matthias Mehlhose, Slawomir Stanczak}
	\IEEEauthorblockA{\textit{Network Information Theory Group, Technische Universität Berlin}, Berlin, Germany \\
		\{m.attar, slawomir.stanczak\}@tu-berlin.de}
	\IEEEauthorblockA{\textit{5G Wirelesss Research Group, Lenovo Deutschland GmbH}, Germany}
	\IEEEauthorblockA{\textit{Fraunhofer Institute for Telecommunications, Heinrich Hertz Institute (HHI)}, Berlin, Germany \\
		\{kaixin.chang, ramez.askar, matthias.mehlhose, slawomir.stanczak\}@hhi.fraunhofer.de}
}


\maketitle

\begin{abstract}
This paper presents a kernel-based adaptive filter that is applied for the digital domain self-interference cancellation (SIC) in a transceiver operating in full-duplex (FD) mode. In FD, the benefit of simultaneous transmission and receiving of signals comes at the price of strong self-interference (SI). In this work, we are primarily interested in suppressing the SI using an adaptive filter namely adaptive projected subgradient method (APSM) in a reproducing kernel Hilbert space (RKHS) of functions. Using the projection concept as a powerful tool, APSM is used to model and consequently remove the SI. A low-complexity and fast-tracking algorithm is provided taking advantage of parallel projections as well as the kernel trick in RKHS. The performance of the proposed method is evaluated on real measurement data. The method illustrates the good performance of the proposed adaptive filter, compared to the known popular benchmarks. They demonstrate that the kernel-based algorithm achieves a favorable level of digital SIC while enabling parallel computation-based implementation within a rich and nonlinear function space, thanks to the employed adaptive filtering method.
\end{abstract}


\section{Introduction}
Full-duplex (FD) communications is a promising technology for future wireless communications with the potential of improved spectral efficiency, reduced end-to-end latency, and higher information secrecy \cite{bharadia_fullDuplex, sabharwal_inbandFullDuplex, taghizadeh_environmentAware, taghizadeh_FDamplifyForward}. In FD radios, information is transmitted and received simultaneously in the same frequency band, offering a much higher capacity to cope with the increasing demand for data transmission. However, these benefits come at the expense of introducing a strong self-interference (SI) signal at the receiver generated by its own transmitter \cite{bharadia_fullDuplex, sabharwal_inbandFullDuplex, Askar2021}.

The SI signal could be canceled in various stages, including propagation domain (passive), relying on the passive isolation among the transmit and receiver front ends, e.g., \cite{everett_passiveSIsuppression}, analog domain (active), e.g., employing auxiliary RF SIC circuitry \cite{askar_activeSIC}, and digital domain (baseband signal processing) \cite{sabharwal_inbandFullDuplex}. The residual SI signal in the digital domain consists of linear and nonlinear parts. The nonlinearity comes from hardware impairments of active components in the RF chain such as power amplifiers and mixers. The issue of digital SI cancellation (DSIC) and its importance have been investigated in recent studies \cite{korpi_adaptiveNonlinear,Li_anAugmentedNonlinear,soriano_adaptiveSelfInterference,anttila_fullDuplexing,kurzo_hardwareImplementation}. DSIC can be classified as a regression problem where the existing residual SI is to be estimated utilizing the knowledge of transmitted signal history and previous instances of the received SI samples.

One of the main challenges in the concept of estimation or regression is the computation of received signal statistics, including spatial and temporal correlation matrices. One solution is to use the law of large numbers, stating that the average of a large number of samples tends toward the expected value as more samples are used. Applying this theorem restricts us to offline techniques. In offline techniques, we require enough time to collect sufficient samples of signals, which is also called batch learning in some contexts. However, it is not desirable in real-time applications where continuous analysis and processing of data is required as data samples arrive sequentially. For example in \cite{kurzo_hardwareImplementation}, it is shown that neural networks are highly capable of learning the hardware nonlinearity. Nonetheless, they are not suitable for dynamic wireless communication systems where a mismatch between the assumed signal statistics for training and the actual scenario leads to significant performance degradation.

As a desirable solution,  benefiting both real-time and low computational complexity and tracking the system changes, adaptive processing has been introduced and received a great deal of attention \cite{theodoridis_machineLearning}. In adaptive filtering, the filter coefficients are updated recursively to decrease predefined error criteria in each iteration. Adaptive filters play a major role in nonstationary environments such as wireless channels, where the filter coefficients update in an online fashion to track alterations. Thus far, several contributions have been made to the DSIC problem in a time-adaptive setting \cite{korpi_adaptiveNonlinear,Li_anAugmentedNonlinear,soriano_adaptiveSelfInterference,anttila_fullDuplexing}.

The new communication links with wide bandwidth, such as millimeter wave wireless channels, operate at high data rates. As a result, DSIC results in high computational complexity as well as high latency. Furthermore, wireless channels are highly dynamic and change rapidly. Hence, we require an adaptive and low-complexity filtering method capable of tracking the changes. Moreover, the key to a successful DSIC is to cope with various forms of transceiver nonlinearity.

Although several adaptive methods have been presented in this context, to the best of our knowledge, there is no detailed investigation on kernel-based methods in reproducing kernel Hilbert spaces (RKHSs). Thanks to the kernel trick\cite{theodoridis_machineLearning}, they provide us with low complexity computations while solving the regression problem in a higher dimensional space. This property enables us to model the linear and nonlinear components of our system using an appropriate kernel.

In this work, we form a kernel by a linear combination of linear and Gaussian kernels. We also adopt the adaptive projected subgradient method (APSM) \cite{theodoridis_adaptiveLearning}, as a nonlinear adaptive filter using the new kernel. One of the most appealing features of APSM is the concept of concurrent processing. It enables us to take full advantage of parallel processing to reduce latency. We demonstrate that the proposed method not only could suppress linear and nonlinear parts of residual SI, but it could also be parallelized.


\section{System Model}
\begin{figure}[tbp]
	\centering
	\resizebox{\columnwidth}{!}{
		\begin{tikzpicture}[node distance=2cm, scale = 1]
			\node (node1) [block1, xshift=7.5cm, yshift=2.0cm, minimum height=1.5cm, line width=1pt, font=\large] {RF chain};
			\node (node2) [block1, right of=node1, xshift=1.5cm, yshift=-2cm, minimum width=1.5cm, minimum height=5cm, line width=1pt, font=\large] {Analog SIC};
			\node (node3) [block1, below of=node1, minimum height=1.5cm, yshift=-2.0cm, xshift=0.0cm, line width=1pt, font=\large] {RF chain};
			\node (node4) [block1, left of=node1, xshift=-1.4cm, yshift=-2.0cm, line width=1pt, font=\large] {$ \hat{f}(\cdot) $};
			\node (node5) [block3, left of=node3, xshift=-4cm, font=\large] {Received signal + residual self-interference};
			\node (node6) [block3, left of=node1, xshift=-4cm, font=\large] {Transmitted data};
			\node (node7) [sum, below of=node4, xshift=0.0cm, yshift=0cm, font=\large] {+};
			\node (node8) [block3, below of=node3, xshift=-3.3cm, font=\large] {Digital SIC};
			\node (node9) [text=black!30!blue, block3, below of=node8, yshift=10cm, font=\large] {Digital domain};
			\node (node10)[mark size=5pt, rotate=180, line width=1pt] (T) at (13.5cm,4.05) {\pgfuseplotmark{triangle}};
			\node (node11)[mark size=5pt, rotate=180, line width=1pt] at (13.5cm,-0.05) {\pgfuseplotmark{triangle}};

			\draw [arrow, line width=2pt] (node4) -- node[anchor=west, yshift=-0.25cm, font=\large] {$-$} (node7);
			\draw [arrow, line width=2pt] (node6) -- (node1);
			\draw [arrow, line width=2pt] (node3) -- node[anchor=north, xshift=-0.6cm, yshift=0.5cm, font=\large] {$+$} (node7);
			\draw [arrow, line width=2pt] (node1) -- (9.85,2);
			\draw [arrow, line width=2pt] (9.85,-2) -- (node3);
			\draw [arrow, line width=2pt] (4.1,2) -- (node4);
			\draw [arrow, line width=2pt] (node7) -- (node5);
			
			\draw [arrow, line width=3pt, color=black!60!green] (13.7,4.1) -- (15.5,4.1);
			\draw [arrow, line width=1pt, color=black!60!green] (15.5,-0.2) -- (13.6,-0.2);
			\draw [thick, line width=3pt, color=black!30!red, ->] (13.7,4) arc (30:-30:4);
			\draw [thick, line width=3pt, color=black!30!red, ->] (8.55,1.8) arc (25:-25:4);
			
			\draw [line width=2pt] (12.15,2) -| (13.5cm,3.9);
			\draw [line width=2pt] (12.15,-2) -| (13.5cm,-0.2);
			
			\draw [dashed, line width=2pt] (3.2,3) -- (5.2,3) -- (5.2,-3) -- (3.2,-3) -- (3.2,3);
			\draw [dashed, black!30!blue, line width=2pt] (6.2,4.5) -- (6.2,-4);
			
			\node [block4, font=\large] at (5.6cm,2.5) {$x[n]$};
			\node [block4, font=\large] at (5.6cm,-1.5) {$y[n]$};
			\node [block4, font=\large] at (13cm,2.5) {Tx};
			\node [block4, font=\large] at (13cm,-1.5) {Rx};
			\node [block5, font=\large] at (14.7,2) {SI};
			\node [block5, font=\large] at (9.35,0) {SI};

			\draw (17,0) --++ (90:5cm);
			\fill [pattern=north west lines] (17,0) rectangle ++ (0.5,5);
			\draw [line width=1pt, color=black!30!red] (16,2.5) -- (17,2);
			\draw [arrow, line width=1pt, color=black!30!red] (17,2) -- (14,0.1);
			\node [block5, font=\large] at (16,0.8) {SI};
		\end{tikzpicture}
	}
	\caption{General schematic of self-interference cancellation.}
	\label{sic}
\end{figure}
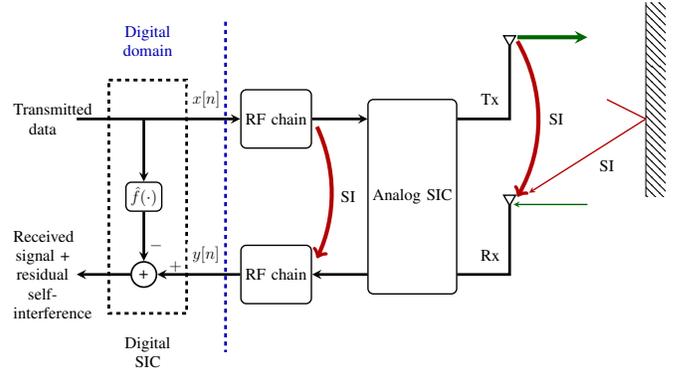

Fig. \ref{sic} illustrates the block diagram of a typical FD system comprising of active analog and digital SI cancellation modules. The transmitted complex signal at time instant $ n $ is denoted by $ x[n] $, while $ y[n] $ represents the digital received signal corrupted by additive white Gaussian noise $ z[n] $. Hence, the nonlinear residual SI signal in the complex digital baseband domain can be written as
\begin{align}
	y[n] = & f(x[n+M_{pre}], \cdots, x[n], \cdots, x[n-M_{post}]) + z[n] \nonumber \\
	= & f(\boldsymbol{x}[n]) + z[n],
	\label{general}
\end{align}
where $ M := M_{pre} + M_{post} + 1 $ is the memory length of the effective SI channel and $ \boldsymbol{x}[n] = \left[ x[n+M_{pre}], \cdots, x[n], \cdots, x[n-M_{post}] \right]^T $, where $ (.)^T $ represents matrix transpose. It should be noted that we have deployed a short time lag of $ M_{pre} $ samples as in \cite{korpi_adaptiveNonlinear} to model the SI channel accurately. Using linear and nonlinear basis functions, the function $ f(\cdot) $ in \eqref{general} may be written as
\vspace{-0.07cm}
\begin{equation}
	\vspace{-0.05cm}
	f(\boldsymbol{x}[n]) \simeq \sum_{k=1}^{K} h_k \psi_k(\boldsymbol{x}[n]),
	\label{nonlinear}
\end{equation}
where $ K $ represents the cardinality of the set of basis functions, and $ h_k $'s denote their corresponding coefficients. In \eqref{nonlinear}, $ \psi_k(\cdot) $ is the $k$-th basis function such as a linear, polynomial, or any other nonlinear function. Our goal is to find the best estimate of $ f(\cdot) $ such that it minimizes the difference
\begin{equation*}
	e[n] = y[n] - \hat{f}(\boldsymbol{x}[n]) \simeq z[n].
\end{equation*}

The main objective of digital SI cancellation is to find a model that provides the best approximation of the system. Indeed, this is a system identification problem.
To our end, for each pair of training data points $ \left( \boldsymbol{x}[n], y[n] \right) \in \mathbb{C}^{M} \times \mathbb{C} $, we need to find a function $ \hat{f}: \mathbb{C}^{M} \rightarrow \mathbb{C} $ such that
\begin{equation}
	| e[n] | = | y[n] - \hat{f}(\boldsymbol{x}[n]) | \leq \varepsilon, \quad n=1,2,\cdots, N,
\end{equation}
where $ \varepsilon \geq 0 $ is a small predefined error tolerance, and $ N $ denotes the total number of observations. We would like to solve this problem using projection-based methods. These are powerful methods in machine learning.

\section{Kernel-Based APSM}
Without loss of generality, we restrict ourselves to real Hilbert spaces. A Hilbert space is a linear vector space like the Euclidean space but with the possibility of having infinite dimensions. The function we estimate can be considered as a point in this Hilbert space. Every Hilbert space $ \mathcal{H} $ is equipped with an inner product operation $ \langle \cdot,\cdot \rangle_{\mathcal{H}} $ and its induced norm. The inner product is a function that assigns a real value to every pair of elements in the space $ ( \mathcal{H} \times \mathcal{H} \rightarrow \mathbb{R} ) $. Accordingly, the induced norm is defined as $ \| \cdot \|^{1/2}_\mathcal{H} = \langle \cdot, \cdot \rangle_{\mathcal{H}} $.

A Hilbert space is called an RKHS if there exists a so-called kernel function $ \kappa (\cdot, \cdot): \mathbb{R}^L \times \mathbb{R}^L \longrightarrow \mathbb{R} $, which is symmetric $ \kappa (\boldsymbol{u}, \boldsymbol{v}) = \kappa (\boldsymbol{v}, \boldsymbol{u}), \quad \forall \boldsymbol{u}, \boldsymbol{v} \in \mathbb{R}^L $, and positive definite, i.e., $ \sum_{n=1}^{N} \sum_{m=1}^{N} a_n a_m \kappa(\boldsymbol{u}_n, \boldsymbol{u}_m) \geq 0, \quad \forall \boldsymbol{u}_1, \cdots, \boldsymbol{u}_N \in \mathbb{R}^L \text{ and } \forall a_1, \cdots, a_N \in \mathbb{R} $. It also has the following properties:
\begin{itemize}
	\item Representation property: $ \kappa(\cdot, \boldsymbol{u}) \in \mathcal{H}, \ \forall \boldsymbol{u} \in \mathbb{R}^L $,
	\item Reproducing property: $ f(\boldsymbol{u}) = \langle f, \kappa(\cdot, \boldsymbol{u}) \rangle_{\mathcal{H}}, \ \forall f \in \mathcal{H}, \ \forall \boldsymbol{u} \in \mathbb{R}^L $.
\end{itemize}

Adopting a kernel enables us to define a mapping from the low-dimensional input space (of size $ L $) to a high-dimensional (possibly infinite) feature space where our problem can be solved linearly, i.e., $	\boldsymbol{u} \in \mathbb{R}^L \longrightarrow Q(\boldsymbol{u}) := \kappa (\cdot, \boldsymbol{u}) \in \mathcal{H}. $ To obtain a solution, it is required to compute the inner products between mapped points efficiently. The well-known reproducing property enables us to replace the inner product operation $ \langle \kappa (\boldsymbol{u}, \cdot), \kappa (\cdot, \boldsymbol{v}) \rangle_{\mathcal{H}} $ with the value of function $ \kappa (\boldsymbol{u}, \cdot) \in \mathcal{H} $ at point $ \boldsymbol{v} \in \mathbb{R}^L $. This leads us to the kernel trick as $\langle \kappa (\boldsymbol{u}, \cdot), \kappa (\boldsymbol{v}, \cdot) \rangle_{\mathcal{H}} = \kappa (\boldsymbol{u}, \boldsymbol{v}), $ decreasing the computational complexity. The kernel trick means that we do not require to compute the inner product in the high-dimensional RKHS, but replace it with the kernel function value in the original input space with much lower complexity. Now, the initial nonlinear problem is cast as a linear one in the RKHS. 

\subsection{SI Cancellation using APSM}
Here our goal is to find a function $ f_n: \mathbb{R}^{L} \rightarrow \mathbb{R} \in \mathcal{H} $ in a time adaptive setting such that $ f_n \in \bigcap_{j \in \mathcal{J}_n} S_j $, where
\begin{equation}
	S_j := \left\{ f \in \mathcal{H} : | f(\boldsymbol{x}_j) - y_j | \leq \varepsilon \right\},
	\label{intersection}
\end{equation}
and $ \boldsymbol{x}_j = \Big[ \mathrm{Re}\{\boldsymbol{x}^T[j]\}, \ \mathrm{Im}\{\boldsymbol{x}^T[j]\} \Big]^T $.
It should be noted that we estimate two distinct functions for real and imaginary parts of $ y[j] $ to deal with I/Q imbalances more effectively. Therefore, $ y_j = \mathrm{Re}\{y[j]\} $ to estimate the real part of the interference signal and accordingly for the imaginary part $ y_j = \mathrm{Im}\{y[j]\} $, i.e., $ y[j] = f_{real}(\boldsymbol{x}_j) + j f_{imag}(\boldsymbol{x}_j) $. The desired functions lie in the intersection of subset $ \mathcal{J}_n $ of nonempty convex sets $ S_j $'s defined by training data points. 

Using the reproducing property of RKHSs, we can rewrite \eqref{intersection} as $ S_j = \left\{ f \in \mathcal{H} : | \langle f, \kappa (\boldsymbol{x}_j, \cdot) \rangle_{\mathcal{H}} - y_j | \leq \varepsilon \right\}, $ which represents hyperslabs. A hyperslab with width $ \varepsilon $ as illustrated in \cite[Fig.~12]{theodoridis_adaptiveLearning} is a convex set. To reach the desired function, we require to project our current estimate to the convex sets described by triple $ (\boldsymbol{x}_j, y_j, \varepsilon) $. The metric projection of point $ f $ onto a nonempty closed convex set $ C $ is expressed as $ P_C(f) = \arg \underset{g \in C}{\min} \| f - g \|_\mathcal{H}. $ The metric projection of our current estimate $ f_n $ onto hyperslab $ S_j $ defined by $ (\boldsymbol{x}_j, y_j, \varepsilon) $ is given by
\vspace{-0.05cm}
\begin{equation}
	P_{S_j}(f_n) = f_n + \beta_j^n \kappa (\boldsymbol{x}_j, \cdot),
	\label{proj}
\end{equation}
\vspace{-0.2cm}where
\begin{equation*}
	\beta_j^n =
	\begin{cases}
		\frac{y_j - \langle f_n, \kappa (\boldsymbol{x}_j, \cdot) \rangle_{\mathcal{H}} - \varepsilon}{\kappa (\boldsymbol{x}_j, \boldsymbol{x}_j)}, & y_j - \langle f_n, \kappa (\boldsymbol{x}_j, \cdot) \rangle_{\mathcal{H}} > \varepsilon,\\
		0, & | y_j - \langle f_n, \kappa (\boldsymbol{x}_j, \cdot) \rangle_{\mathcal{H}} | \leq \varepsilon,\\
		\frac{y_j - \langle f_n, \kappa (\boldsymbol{x}_j, \cdot) \rangle_{\mathcal{H}} + \varepsilon}{k (\boldsymbol{x}_j, \boldsymbol{x}_j)}, & y_j - \langle f_n, \kappa (\boldsymbol{x}_j, \cdot) \rangle_{\mathcal{H}} < -\varepsilon.
	\end{cases}
\end{equation*}

In our problem, every training data point $ (\boldsymbol{x}_n, y_n) $ arrives sequentially defining a new hyperslab. We look for a function that resides in their intersections. However, it is not possible to consider all hyperslabs due to computational restrictions and the dynamic nature of wireless communication systems where the intersection might become empty. Therefore, we restrict ourselves to a subset $ \mathcal{J}_n \subset \{1,2,\cdots,n\} $ of indices with cardinality $ q = | \mathcal{J}_n | $. A typical choice for $ \mathcal{J}_n $ in a dynamic system could be $ q $ most recent samples. To our goal, we apply the APSM algorithm capable of concurrent processing to reduce latency.

The APSM algorithm operates as follows:
\begin{itemize}
	\item The algorithm starts with time index $ n=0 $, a predefined positive value $ q $, and a positive $ \varepsilon $.
	\item Choose an arbitrary initial point $ f_0 \in \mathcal{H} $.
	\item Compute the projections and the parameter $ \mathcal{M}_n $ as
	\begin{equation*}
		\resizebox{0.9\hsize}{!}{$
			\mathcal{M}_n :=
			\begin{cases}
				\frac{\sum_{j \in \mathcal{J}_n} \omega_j^n \| P_{S_j} (f_n) - f_n \|^2_\mathcal{H}}{\| \sum_{j \in \mathcal{J}_n} \omega_j^n P_{S_j} (f_n) - f_n \|_\mathcal{H}^2}, & \text{if } f_n \neq \sum_{j \in \mathcal{J}_n} \omega_j^n P_{S_j} (f_n)\\
				1, & \text{otherwise}.
			\end{cases}
		$}
	\end{equation*}
	\item For each training point $ (\boldsymbol{x}_n, y_n) $, compute the next estimate as $ f_{n+1} = f_n + \mu \mathcal{M}_n \left( \sum_{j \in \mathcal{J}_n}^{} \omega_j^n P_{S_j} (f_n) - f_n \right), $	where $ \sum_{j \in \mathcal{J}_n} \omega_j^n = 1 $ and $ \omega_j^n \geq 0 $.
\end{itemize}
It can easily be verified by induction \cite{theodoridis_adaptiveLearning} that starting from an initial point $ f_0 $, the adaptive filter $ f_n $ is of the general form
\vspace{0cm}
\begin{equation}
	f_n(\cdot) = \sum_{j=0}^{n-1} \gamma_j^{(n)} \kappa (\boldsymbol{x}_j, \cdot),
	\label{representer}
\end{equation}
where $ \gamma_j^{(n)} $'s are the coefficients to be updated at each time.

\subsection{Various Kernels for APSM}
As discussed earlier, a kernel implicitly maps the elements in the input space to a possibly higher dimensional feature space. Given the structure of our data, choosing the right kernel plays a vital role in the performance of DSIC. An informed choice is the linear kernel $ \kappa_L (\boldsymbol{u}, \boldsymbol{v}) = \boldsymbol{u}^T \boldsymbol{v}, \ \forall \boldsymbol{u}, \boldsymbol{v} \in \mathbb{R}^L $, inasmuch as the SI signal is comprised of strong linear part. However, we know that nonlinearity is also a significant part of the SI signal. A celebrated nonlinear kernel forming a RKHS is the Gaussian function, $	\kappa_G (\boldsymbol{u}, \boldsymbol{v}) = \exp \left( -\xi (\boldsymbol{u}-\boldsymbol{v})^T (\boldsymbol{u}-\boldsymbol{v}) \right), \ \forall \boldsymbol{u}, \boldsymbol{v} \in \mathbb{R}^L, $ where $ \xi $ represents the kernel width and takes a positive real value. For the Gaussian kernel, the induced RKHS is of infinite dimension.

It is important to choose a proper kernel and its associate RKHS representing the signal characteristics. Hence, we search for a function in a bigger Hilbert space including both linear and nonlinear functions. We apply a weighted sum of linear and Gaussian kernels as a new kernel, $ \kappa (\boldsymbol{u}, \boldsymbol{v}) = w_L \kappa_L (\boldsymbol{u}, \boldsymbol{v}) + w_G \kappa_G (\boldsymbol{u}, \boldsymbol{v}), \quad \forall w_L, w_G > 0. $ The sum space of RKHSs associated with the linear and Gaussian kernels is defined as $ \mathcal{H}^+ = \left\lbrace f=f_L+f_G | f_L \in \mathcal{H}_L, \ f_G \in \mathcal{H}_G \right\rbrace, $ where the inner product in this Hilbert space is presented by $	\langle f, g \rangle_{\mathcal{H}^+,w} = w_L^{-1} \langle f_L, g_L \rangle_{\mathcal{H}_L} + w_G^{-1} \langle f_G, g_G \rangle_{\mathcal{H}_G} $ \cite{yukawa_onlineLearning}.

\begin{figure*}[!t]
	\centering
	\includegraphics[width=1.71\columnwidth]{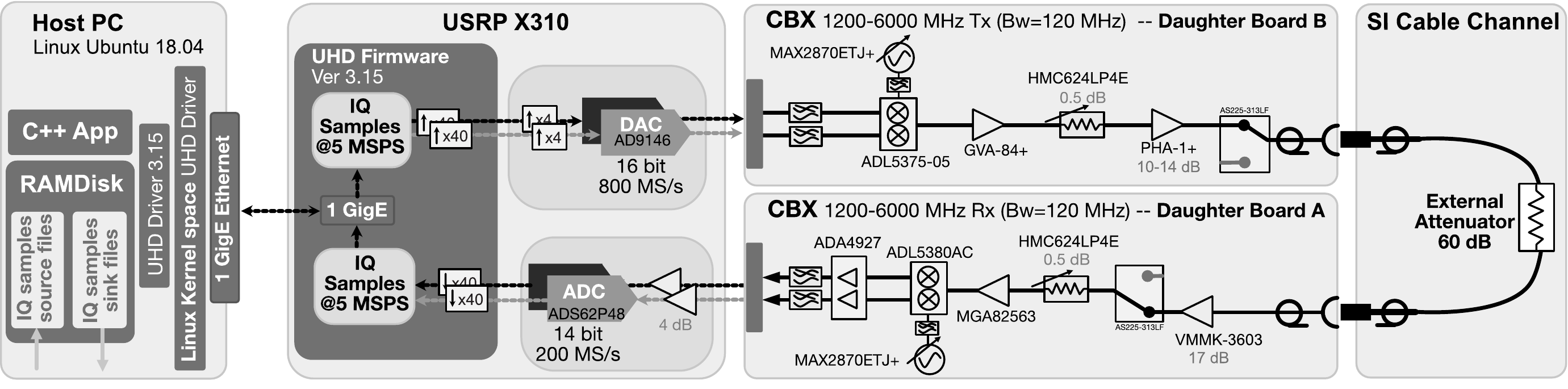}
	\vspace{-0.25cm}
	\caption{Hardware-in-the-loop experimental setup -- the setup comprises a USRP X310, two CBX-120 transceiver front-end daughterboards and a host PC.}
	\vspace{-0.4cm}
	\label{testbed}
\end{figure*}

\subsection{Sparsification (Dictionary Learning)}\label{dict}
Working in an RKHS provides us with a typically higher dimensional space in which we can solve our problem linearly. The chosen kernel function determines the dimensionality of the induced RKHS. For example, the linear kernel $ \kappa_L (\boldsymbol{u}, \boldsymbol{v}) $ gives us an RKHS of constant dimension. However, it is known that the RKHS associated with the Gaussian kernel $ \kappa_G (\boldsymbol{u}, \boldsymbol{v}) $ is of infinite dimension \cite{theodoridis_adaptiveLearning}. This property causes practical issues when applying the APSM algorithm.

Considering \eqref{proj} and \eqref{representer}, it can be seen that the adaptive function grows constantly in that the number of coefficients and bases $ \kappa (\boldsymbol{x}_j, \cdot) $ increases as a new training point arrives. Not only do we require to keep them in the memory, they could also prove the computations prohibitively expensive as iterations evolve. It should be emphasized that at each time instant, we search for a solution in a subspace of $ \mathcal{H} $ spanned by bases $ \kappa (\boldsymbol{x}_j, \cdot), \ j=0,1,\cdots,n $.

To comply with memory and computation requirements, we adopt a mechanism called sparsification, which builds a dictionary. The so-called dictionary at time instant $ n $, denoted by $ \mathcal{D}_n $, is a set of basis functions by which the solution space is spanned, i.e., $ \mathcal{D}_n := \{ \kappa (\boldsymbol{x}_b, \cdot) \}_{b=1}^{B} $. Indeed, the solution in the space $ \mathcal{H} $ is a linear combination of elements, also called atoms, in the dictionary. Its cardinality $ B $ specifies the size of the linear subspace, $ V_n := \text{span}\{\mathcal{D}_n\} $. To construct a rich dictionary, it is sensible to learn $ \mathcal{D}_n $ based on novelty criteria. One popular novelty criterion is the approximate linear dependency (ALD) \cite{slavakis_slidingWindow, engel_kernelRLS}. According to ALD, only those elements that are approximately linearly independent of the existing ones enter the dictionary. When a new element $ \kappa (\boldsymbol{x}_{n+1}, \cdot) $ arrives, its orthogonal projection to the subspace $V_n$ is calculated. If the distance of the element from its projection is larger than a positive threshold $ \alpha $, it implies that it contains novel information and enters the dictionary. Mathematically, the above can be formulated as
\begin{equation}
	\| \kappa (\boldsymbol{x}_{n+1}, \cdot) - P_{V_n} \left( \kappa (\boldsymbol{x}_{n+1}, \cdot) \right) \|_\mathcal{H} \geq \alpha.
	\label{ald}
\end{equation}
It should be noted that we employ only one dictionary for both real and imaginary functions being estimated.

\section{Hardware-in-the-Loop Experiment}

In this section, we report the hardware-in-the-loop experiment in which we evaluated the performance of the proposed kernel-based APSM for DSIC in a FD communication scenario with offline digital samples. The performance of the considered methods is compared in terms of mean squared error (MSE), $ \text{MSE}_\text{dB} = 10 \log_{10} \left( \frac{\sum_{n=0}^{N-1}{|y[n] - \hat{f}(\boldsymbol{x}[n])|^2}}{N} \right) $, 
as well as the rate of convergence. We conducted a number of $ 100 $ independent experiments. Then, performance values and curves were calculated by taking uniform averages.

\begin{table}[bp]
	\caption{The measurement parameters}
	\begin{center}
		\begin{tabular}{|c||c|}
			\hline
			\textbf{Parameter} & \textbf{Value} \\
			\hline
			Carrier frequency & $ 2.4 $ GHz \\
			\hline
			Signal (analog) bandwidth & $ 5 $ MHz \\
			\hline
			Transmitter and receiver sampling rate & $ 200 $/$ 800 $ MS/s \\
			\hline
			Number of training realizations & $ 100 $ \\
			\hline
			Number of test samples & $ 500,\!000 $ \\
			\hline
			$ M_{pre} $ and $ M_{post} $ & $ 10 $ \\
			\hline
		\end{tabular}
	\vspace{-0.2cm}
	\label{tab1}
	\end{center}
\end{table}

\begin{figure}[bp]
	\centering
	\vspace{-0.3cm}
	\includegraphics[width=0.65\columnwidth]{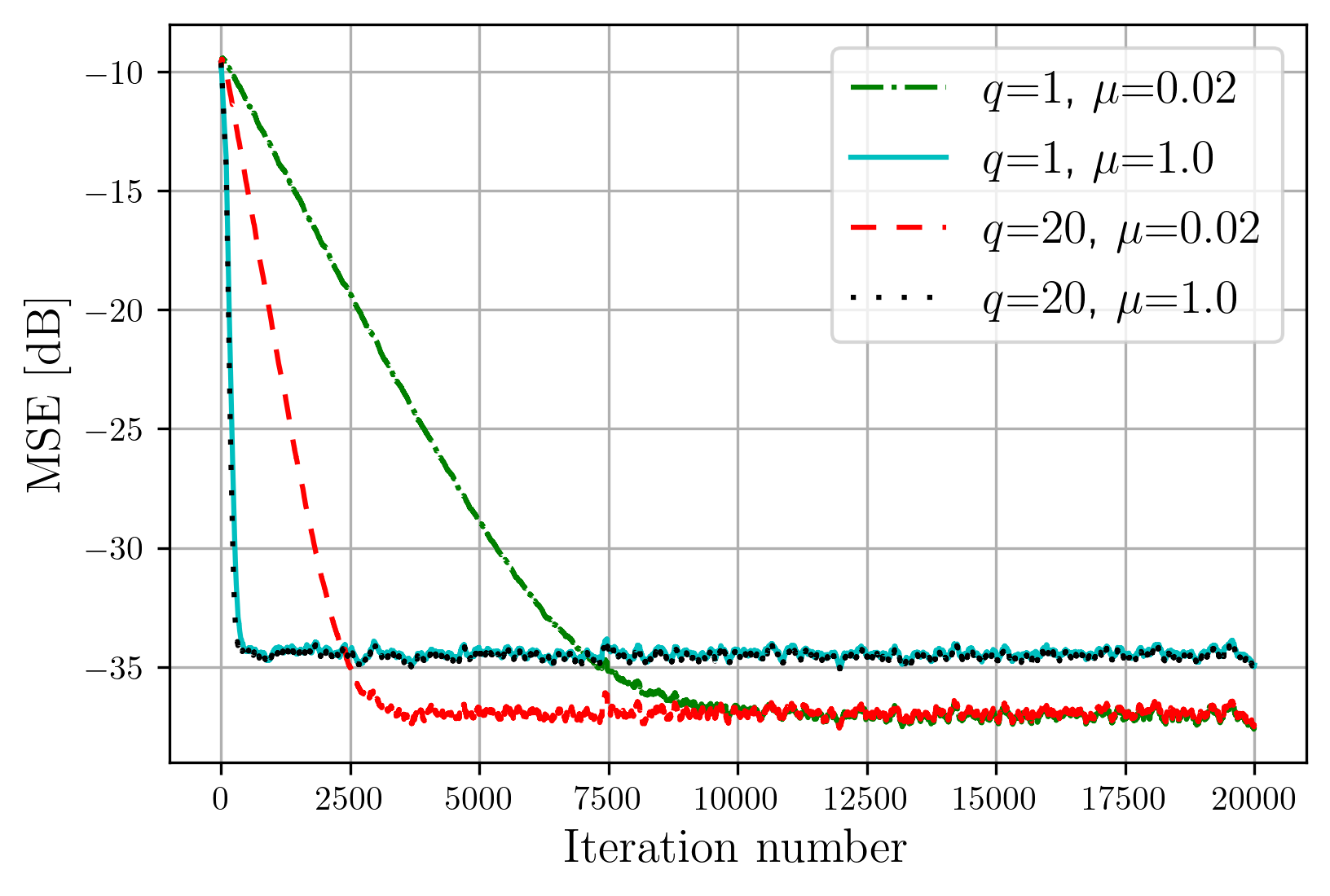}
	\vspace{-0.3cm}
	\caption{Impact of $\mu$ and $q$ on the performance and speed of the APSM filter using a linear kernel when $\varepsilon=0.001$.}
	\label{mu}
\end{figure}

The experimental setup was an FD transceiver prototype built from off-the-shelf components. We utilized a USRP X310 software-defined radio (SDR) platform with two CBX-120 front-end daughterboards. A diagram of the experimental setup is shown in Fig. \ref{testbed}. A host PC was used to transmit and capture digital samples -- connected to the USRP via a 1-Gbps Ethernet link. The transmission signal was a sequence of complex Gaussian i.i.d. (random) samples. Table \ref{tab1} summarizes the experiment parameters, including several USRP configuration parameters. To emulate a static single-tap self-interference channel with sufficient RF SI cancellation, we connected the receiver's LNA input directly to the transmitter PA output via a 60-dB RF attenuator. Although a static single-tap channel does not represent a realistic SI channel \cite{Askar2020}, we considered it to perform a preliminary algorithm assessment.

Fig. \ref{mu} compares the performance of APSM using the linear kernel with various values for $ q $ and $ \mu $ by plotting the MSE with respect to iteration number, which corresponds to filter update. The step size $ \mu $ regulates the convergence rate of the algorithm. The good performance of choosing a small value, say $ \mu = 0.02 $, comes at the price of slow convergence for $ q = 1 $. As expected, we reach the steady state faster by increasing $ \mu $ while losing some performance, i.e., $\mu = 1.0$ with $q=1$. However, we could increase the speed by means of concurrent processing (moving forward to the intersection of $ q $ hyperslabs). Indeed by taking advantage of concurrent processing, e.g. $q=20$, we reach convergence fast while choosing a smaller value for $ \mu = 0.02 $ and an apt value for $ \varepsilon $ representing the width of hyperslabs. It should be noted that big values for $ q $ corresponding to high computational complexity is of no use in case of big step sizes, for instance $\mu = 1.0$, as illustrated in Fig. \ref{mu}.
\begin{figure}[tbp]
	\centering
	\includegraphics[width=0.65\columnwidth]{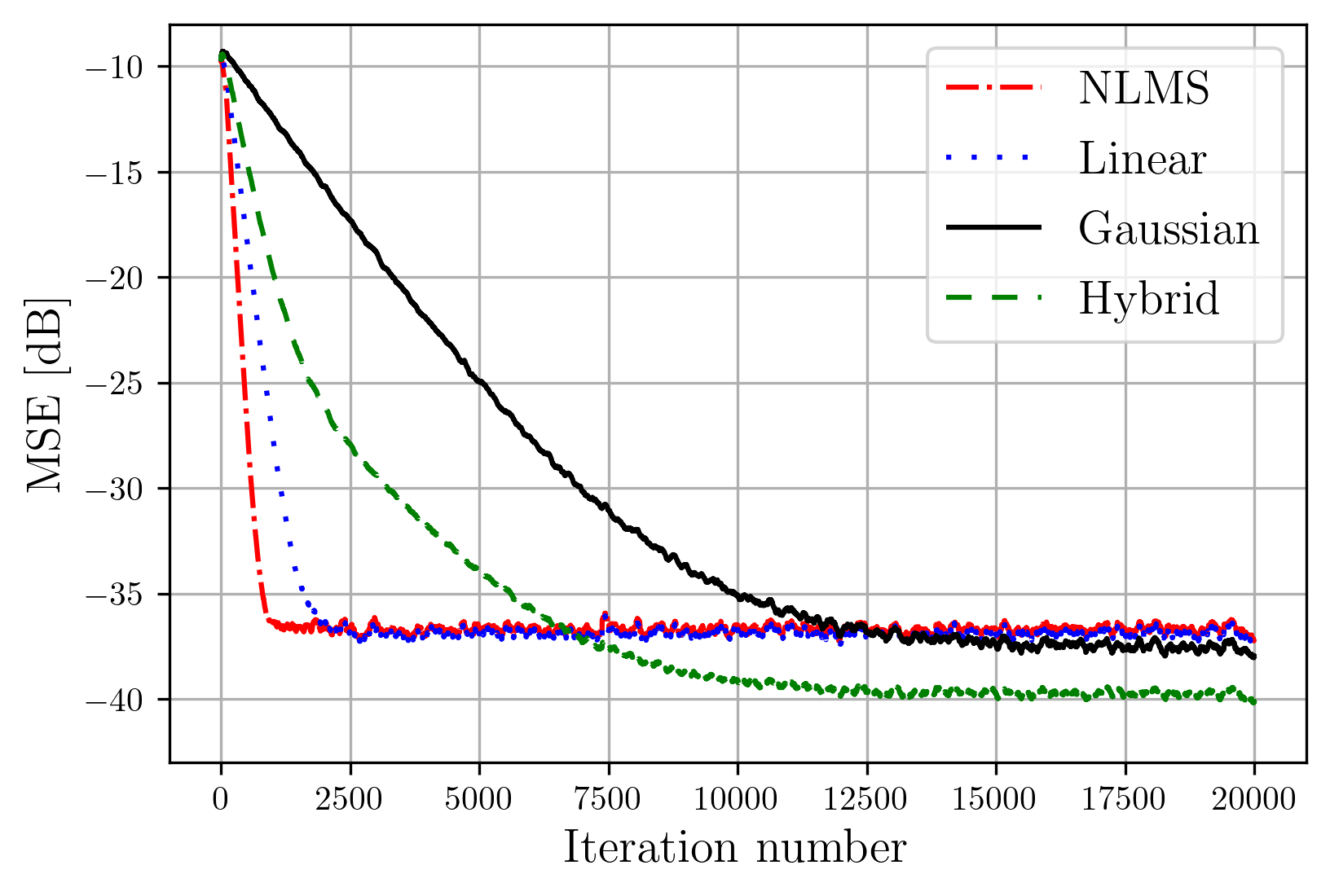}
	\vspace{-0.3cm}
	\caption{Performance comparison of APSM employing various kernels when $\varepsilon=0.001$, $\mu=0.1$, and $q=1$ along with NLMS using the step size of $0.1$.}
	\label{compare}
\end{figure}

Fig. \ref{compare} depicts the performance of the considered APSM filters including linear, Gaussian, and the sum space kernel, referred to as hybrid, along with that of normalized least mean squares (NLMS) filter with respect to iteration number. The hyperparameters of every filter are chosen by cross-validation such that they deliver their best performance in terms of MSE. The width for the Gaussian and hybrid kernels are $ \xi = 0.0715 $ and $ \xi = 0.225 $, respectively. We set $w_L = 0.1$ and $w_G = 0.9$ for the hybrid kernel. It is seen that the linear kernel provides a better convergence rate rather than the Gaussian. Furthermore, we observe that the Gaussian kernel delivers a slightly less MSE (around $ 1 $ dB) than the linear one at the price of being slow. However, by unifying these two kernels and forming a weighted one, we are able to benefit considerably from their advantages. Therefore, we achieve much better performance by the hybrid kernel, which considers both linear and nonlinear components. Not only it converges faster than the Gaussian kernel, but it also delivers the least MSE. We also include the learning curve for NLMS as a reference in Fig. \ref{compare}. As expected, it gives a performance comparable to the linear kernel.

The complexity analysis of the proposed methods in terms of dictionary size for each kernel is listed in Table \ref{tab2}. As mentioned in Section \ref{dict}, the dictionary size when employing the linear kernel is constant, equal to the size of input space $ L $. However, utilizing the Gaussian kernel leads to an RKHS of infinite dimension justifying the importance of dictionary learning. To achieve the best performance, we store all the elements satisfying the condition in \eqref{ald}, where $ \alpha = 0.1 $. As a result, the constructed dictionary becomes considerably large. By contrast, we require a perfectly reasonable number of atoms in the dictionary for the hybrid kernel.

\begin{figure}[tp]
	\centering
	\includegraphics[width=0.65\columnwidth]{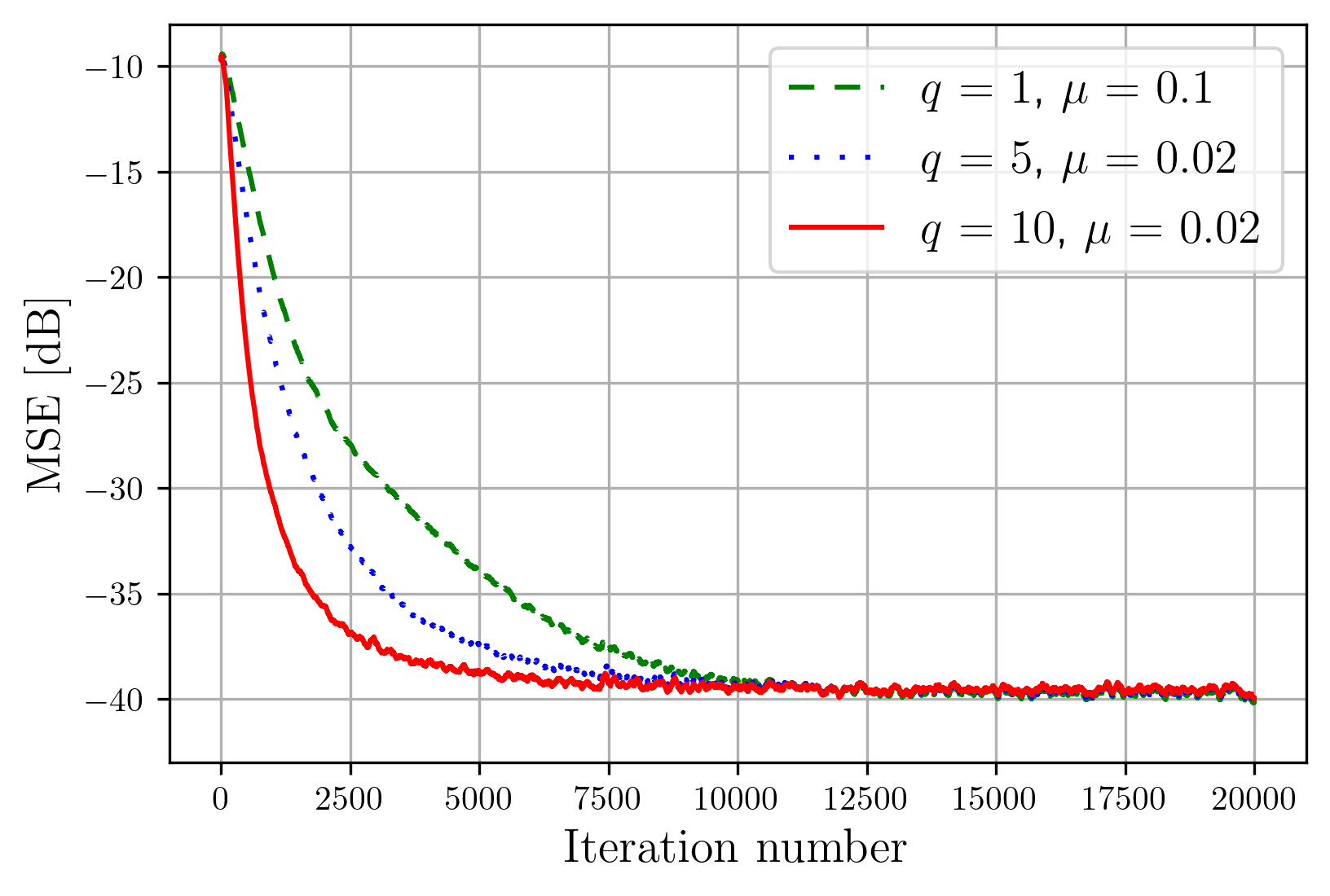}
	\vspace{-0.3cm}
	\caption{Key feature of concurrent processing in APSM with the hybrid kernel when $\varepsilon=0.001$, $ \xi = 0.225 $, $w_L = 0.1$, and $w_G = 0.9$.}
	\label{hybrid}
\end{figure}

Lastly, note once again that concurrent processing is one of the most appealing features of APSM. It enables us to take full advantage of parallel processing to reduce latency (converging faster) by increasing hyperparameter $q$. Fig. \ref{hybrid} illustrates the impact of this key feature in case of hybrid kernel. As mentioned earlier, we can adopt smaller values for step size $ \mu $ while increasing $ q $.

\section{Conclusions}
In this work, we integrated the concept of reproducing kernel Hilbert spaces into the context of digital self-interference cancellation in full-duplex communications. The residual self-interference in the digital domain consists of both linear and nonlinear components. By the combination of linear and Gaussian kernels, we formed a new kernel by which our nonlinear problem turns into a linear one in an RKHS of functions. We adopted the adaptive projection subgradient method as a nonlinear adaptive filter to cope with the dynamic changes in a wireless environment. The simulation results demonstrated that the proposed adaptive filter can effectively predict the residual self-interference. It is also capable of performing parallel processing to reduce latency. We also investigated the impact of hyperparameters on the performance. 

\begin{table}[tp]
	\vspace{-0.3cm}
	\caption{Complexity analysis of algorithms in terms of dictionary sizes}
	\begin{center}
		\begin{tabular}{|c||c|}
			\hline
			\textbf{Kernel} & \textbf{Average dictionary size} \\
			\hline
			Linear & $ 42 $ \\
			\hline
			Gaussian & $ 1188.2 $ \\
			\hline
			Hybrid & $ 42 + 35.8 $ \\
			\hline
		\end{tabular}
		\label{tab2}
	\end{center}
	\vspace{-0.4cm}
\end{table}

\section*{Acknowledgment}
This work was supported in part by the DFG project DUPLINK under grant STA 864/12-1 and the BMBF project OTB-5G+ under funding 16KIS0985.

\bibliographystyle{IEEEtran}
\bibliography{references}

\begin{thebibliography}{10}
\providecommand{\url}[1]{#1}
\csname url@samestyle\endcsname
\providecommand{\newblock}{\relax}
\providecommand{\bibinfo}[2]{#2}
\providecommand{\BIBentrySTDinterwordspacing}{\spaceskip=0pt\relax}
\providecommand{\BIBentryALTinterwordstretchfactor}{4}
\providecommand{\BIBentryALTinterwordspacing}{\spaceskip=\fontdimen2\font plus
\BIBentryALTinterwordstretchfactor\fontdimen3\font minus
  \fontdimen4\font\relax}
\providecommand{\BIBforeignlanguage}[2]{{%
\expandafter\ifx\csname l@#1\endcsname\relax
\typeout{** WARNING: IEEEtran.bst: No hyphenation pattern has been}%
\typeout{** loaded for the language `#1'. Using the pattern for}%
\typeout{** the default language instead.}%
\else
\language=\csname l@#1\endcsname
\fi
#2}}
\providecommand{\BIBdecl}{\relax}
\BIBdecl

\bibitem{bharadia_fullDuplex}
D.~Bharadia, E.~McMilin, and S.~Katti, ``Full duplex radios,'' \emph{SIGCOMM
  Comput. Commun. Rev.}, vol.~43, no.~4, Aug. 2013.

\bibitem{sabharwal_inbandFullDuplex}
A.~Sabharwal, P.~Schniter, D.~Guo, D.~W. Bliss, S.~Rangarajan, and R.~Wichman,
  ``In-band full-duplex wireless: Challenges and opportunities,'' \emph{IEEE J.
  Sel. Areas Commun.}, vol.~32, no.~9, 2014.

\bibitem{taghizadeh_environmentAware}
O.~Taghizadeh, P.~Sirvi, S.~Narasimha, J.~A.~L. Calvo, and R.~Mathar,
  ``Environment-aware minimum-cost wireless backhaul network planning with
  full-duplex links,'' \emph{IEEE Syst. J.}, vol.~13, no.~3, 2019.

\bibitem{taghizadeh_FDamplifyForward}
O.~Taghizadeh, A.~C. Cirik, and R.~Mathar, ``Hardware impairments aware
  transceiver design for full-duplex amplify-and-forward {MIMO} relaying,''
  \emph{IEEE Trans. Wireless Commun.}, vol.~17, no.~3, 2018.

\bibitem{Askar2021}
R.~Askar, J.~Chung, Z.~Guo, H.~Ko, W.~Keusgen, and T.~Haustein, ``Interference
  handling challenges toward full duplex evolution in {5G} and beyond cellular
  networks,'' \emph{IEEE Wireless Commun.}, vol.~28, no.~1, 2021.

\bibitem{everett_passiveSIsuppression}
E.~Everett, A.~Sahai, and A.~Sabharwal, ``Passive self-interference suppression
  for full-duplex infrastructure nodes,'' \emph{IEEE Trans. Wireless Commun.},
  vol.~13, no.~2, 2014.

\bibitem{askar_activeSIC}
R.~Askar, T.~Kaiser, B.~Schubert, T.~Haustein, and W.~Keusgen, ``Active
  self-interference cancellation mechanism for full-duplex wireless
  transceivers,'' in \emph{Int. Conf. Cognitive Radio Oriented Wireless Netw.
  and Commun.}, 2014.

\bibitem{korpi_adaptiveNonlinear}
D.~Korpi, Y.-S. Choi, T.~Huusari, L.~Anttila, S.~Talwar, and M.~Valkama,
  ``Adaptive nonlinear digital self-interference cancellation for mobile inband
  full-duplex radio: Algorithms and {RF} measurements,'' in \emph{IEEE Global
  Commun. Conf.}, 2015.

\bibitem{Li_anAugmentedNonlinear}
Z.~Li, Y.~Xia, W.~Pei, K.~Wang, and D.~P. Mandic, ``An augmented nonlinear
  {LMS} for digital self-interference cancellation in full-duplex
  direct-conversion transceivers,'' \emph{IEEE Trans. Signal Process.},
  vol.~66, no.~15, 2018.

\bibitem{soriano_adaptiveSelfInterference}
F.~J. Soriano-Irigaray, J.~S. Fernandez-Prat, F.~J. Lopez-Martinez,
  E.~Martos-Naya, O.~Cobos-Morales, and J.~T. Entrambasaguas, ``Adaptive
  self-interference cancellation for full duplex radio: Analytical model and
  experimental validation,'' \emph{IEEE Access}, vol.~6, 2018.

\bibitem{anttila_fullDuplexing}
L.~Anttila, V.~Lampu, S.~A. Hassani, P.~P. Campo, D.~Korpi, M.~Turunen,
  S.~Pollin, and M.~Valkama, ``Full-duplexing with {SDR} devices: Algorithms,
  {FPGA} implementation, and real-time results,'' \emph{IEEE Trans. on Wireless
  Commun.}, vol.~20, no.~4, 2021.

\bibitem{kurzo_hardwareImplementation}
Y.~Kurzo, A.~T. Kristensen, A.~Burg, and A.~Balatsoukas-Stimming, ``Hardware
  implementation of neural self-interference cancellation,'' \emph{IEEE Trans.
  Emerg. Sel. Topics Circuits Syst.}, vol.~10, no.~2, 2020.

\bibitem{theodoridis_machineLearning}
S.~Theodoridis, \emph{Machine Learning: A Bayesian and Optimization
  Perspective}, 1st~ed.\hskip 1em plus 0.5em minus 0.4em\relax USA: Academic
  Press, Inc., 2015.

\bibitem{theodoridis_adaptiveLearning}
S.~Theodoridis, K.~Slavakis, and I.~Yamada, ``Adaptive learning in a world of
  projections,'' \emph{IEEE Signal Process. Mag.}, vol.~28, no.~1, 2011.

\bibitem{yukawa_onlineLearning}
M.~Yukawa, ``Online learning based on iterative projections in sum space of
  linear and {G}aussian reproducing kernel {H}ilbert spaces,'' in \emph{IEEE
  Int. Conf. Acoust., Speech and Signal Process.}, 2015.

\bibitem{slavakis_slidingWindow}
K.~Slavakis and S.~Theodoridis, ``Sliding window generalized kernel affine
  projection algorithm using projection mappings,'' \emph{EURASIP J. Adv. in
  Signal Process.}, 2008.

\bibitem{engel_kernelRLS}
Y.~Engel, S.~Mannor, and R.~Meir, ``The kernel recursive least-squares
  algorithm,'' \emph{IEEE Trans. Signal Process.}, vol.~52, no.~8, 2004.

\bibitem{Askar2020}
R.~Askar, M.~Mazhar~Sarmadi, F.~Undi, M.~Peter, W.~Keusgen, and T.~Haustein,
  ``Time dispersion parameters of indoor self-interference radio channels in
  sub-7-{GHz} bands,'' in \emph{IEEE Wireless Commun. Netw. Conf. Workshops},
  2020.

\end{thebibliography}

\end{document}